\documentclass[manuscript]{aastex62}

\received{}
\revised{}
\accepted{}
\shorttitle{An interpretation of variability of PG1302-102}
\shortauthors{Kova{\v c}evi{\' c} et al.}

\begin{document}

\title{The optical variability of Supermassive Black Hole Binary Candidate 
PG1302-102: \\
Periodicity and perturbation in the light curve}

\correspondingauthor{Andjelka B. Kova{\v c}evi{\'c} }
\email{andjelka@matf.bg.ac.rs}

\author{Andjelka B. Kova{\v c}evi{\'c}}
\affil{Department of astronomy, Faculty of mathematics, University of Belgrade \\
Studentski trg 16, Belgrade, 11000, Serbia}

\author{Luka {\v C}. Popovi{\'c}}
\affiliation{Astronomical observatory Belgrade \\
Volgina 7, P.O.Box 74 11060, Belgrade,  11060, Serbia}
\affiliation{Department of astronomy, Faculty of mathematics, University of Belgrade \\
Studentski trg 16, Belgrade, 11000, Serbia}
\author{Sa{\v s}a  Simi{\'c}}
\affiliation{Faculty of Sciences, Department of Physics, Univerisity of Kragujevac\\
Radoja Domanovica 12, Kragujevac, 34000,  Serbia}

\author{Dragana Ili{\'c}}
\affiliation{Department of astronomy, Faculty of mathematics, University of Belgrade \\
Studentski trg 16, Belgrade, 11000, Serbia}



\begin{abstract}

The photometric light curve of   PG1302-102 shows  periodic variability which makes this object  one of the most plausible supermassive black hole binary candidate.
Interestingly, the most recent   study of its updated optical light curve  reports  a decrease in a significance of periodicity which  may suggest   that the binary model is less favorable.
Here, we  model the   PG 1302-102 light curve, spanning almost 20 years,  with a
 supermassive  black hole binary system in which a perturbation  in the  accretion  disk of more massive component  is present.
Our model reproduces well the observed light curve with a slight perturbation of
a sinusoidal feature, and predicts that a slightly larger period than previously
reported, of about 1899 days,  could  arise due to a  cold spot in the disk of
 more massive component of a close, unequal-mass ($\frac{\mathrm{m}_{1}}{\mathrm{m}_{2}}=0.1$) black hole  binary system. The light curve  resembles the pattern of sinusoid-like shape within a few years, which could be confirmed by future observations.   In addition, using our hybrid method for periodicity detection,  we show that the periods in the  observed ($1972\pm 254$ days) and modeled
($1873 \pm \ 250$ days) light curves are  within one-sigma,  which is also consistent with our physical model prediction and with previous findings. Thus, both  the periodic nature and its slight fluctuation  of the light curve of PG1302-102  is evident from our physical model and confirmed by  the hybrid method for periodicity detection.

\end{abstract}



\section{Introduction}\label{sec:intro}

The hierarchical structure formation model of galaxies suggests su\-per\-mass\-ive black hole binaries (SMBHB) should be common in the galactic nuclei \citep[see recent analysis ][and references therein]{2016ApJ...828...73K,2017MNRAS.464.3131K}; yet these systems are extremely difficult to identify at sub-parsec  separations  even in the local Universe \citep{2017arXiv171202362D}.
In the era of mul\-ti\-mess\-en\-ger astrophysics, the importance of  the SMBHB at sub-parsec distance surpasses  the understanding of evolutionary processes. They are recognized  as targets for associating gravitational waves with el\-ec\-tro\-mag\-ne\-tic counterparts  \citep[][]{2018ApJ...853L..17B}. 
Such a possibility is quite likely,  because merging black holes could interact with:  a cir\-cum\-bi\-na\-ry accretion disk,  remnant gas between the black holes, and  a mag\-ne\-to\-sph\-ere. All these interactions could  contribute to electromagnetic counterparts \citep[see][and references therein]{2010Sci...329..927P}.

As such compact accreting source  can not be resolved spatially, the presence of any periodic signal    should be detected indirectly from  the  SMBHB  effects \citep[see 
e.g.][]{2012NewAR..56...74P,2012ApJ...759..118B,bon16,li16}    either on  the surrounding accreting gas or precessing jet \citep[see][and references therein]{2018MNRAS.476.4617C,2018MNRAS.tmp..975B}.
However,  these systems   exhibit random fluctuations, whose Fourier spectra  follow power law with indices larger than zero, so called  red noise \citep{1978ComAp...7..103P},  making periodicity more difficult to  detect \citep{2010MNRAS.402..307V}.

A particularly  appealing recent case is the quasar PG1302-102 \citep{2015Natur.518...74G}. By comparison with other SMBHB candidates,  PG1302-102  photometric light curve  resembles rather sinusoidal   structure. Still, its  light curve is not strictly  periodic \citep{2015Natur.525..351D}. 

\cite{2015Natur.518...74G}  reports an evidence of a binary  system with a $\sim$ 4yr  rest-frame period based on the analysis of  data from Catalina Real-Time Transient Survey --CRTS. Additionally, \cite{2015ApJ...814L..12J}  and \cite{2017MNRAS.470.1198D}  reports periodic variability of PG1302-102 in the infrared. Up to now, a new clues  to its variability has emerged.
Namely, it seems that  adding recent observations from All-Sky Automated Survey for Supernovae --ASAS-SN, which are analyzed in details by \cite{2018ApJ...859L..12L},  shows that the evidence for periodicity decreases, and that  further new observations would clarify the significance of the
SMBHB model. 
 The first   aim of our work is  to model  the optical  light curve   with a  perturbation in the disk of more massive component  in the  SMBHB  \citep{Popovic18} which slightly perturb the sinusoidal signal and to forecast the  light curve variability in the next  few years. 
The reason for choosing such approach is that the standard SMBHB model assumes that an accretion disk surrounds  at least more  massive black hole and that  the outcoming variability and structural changes are determined by dynamical characteristics of the disk as well as the interaction of the SMBHB--disks system \citep{2005A&A...431..831L}.
 The second  aim  is to test our newly proposed hybrid method for oscillation detection in the light curves of quasars   \citep[which was presented in][]{2018MNRAS.475.2051K}, on  both observed and modeled light curves.

The structure of the paper is as follows.  We first introduce  our physical model in  Section \ref{sec:modelsmbh}. We then present briefly the data and hybrid method for periodicity detection  in  Section \ref{sec:datmet}.  The results are described and  discussed in Section \ref{sec:resdisc}. Summary of our findings concludes our paper in Section \ref{sec:con}.

\section{The model: SMBBHs and perturbation in the emission disk}  \label{sec:modelsmbh}

There are several approaches to model the emission  from  the
SMBHB \citep[see ][and references therein]{2012NewAR..56...74P}.
Here we utilize the model described in  \cite{Simic16}.
 The model   is  able to  include  perturbations in one of the component disks (or both of them)  which are resulting in   either an  amplification or  attenuation of the flux of the system. 
 The model can 
be shortly described as following. Adopted geometry of the SMBHB system  assumes
 two supermassive  black holes (with mass  of a less massive component   $m_{1}$,  and more massive component $m_{2}$,  i.e. $\mathbf{m_{1}<m_{2}}$ and $\mathbf{q=\frac{m_{1}}{m_{2}}}$) which  orbit the  barycenter of the system, in the plane inclined  at an angle $\theta$  with respect to the observer.
Accretion disks around each black holes are coplanar  with  the orbital plane. 

 Both accretion disks  are classical geometrically thin optically thick relativistic disk   proposed by \cite{Shakura73},  which  are thermalized due to the friction of rotating matter and radiate continuum emission in the UV, optical and IR band. 
 The disk effective temperature ($T_{\mathrm{eff}}$)    decreases with the radius $R$,   and is given  with the following  expression adopted from  \cite{2016ASSL..440....1L}:
\begin{equation}
T_{\mathrm{eff}}[K]=2\cdot10^5 \left(\frac{10^8}{m_i}\right)^{1/4} \left(\frac{R_{\mathrm{in}}}{R}\right)^{\frac{3}{4}}\left(1-\sqrt{\frac{R_{\mathrm{in}}}{R}}\right)
\label{eq:TodR}
\end{equation}
where  $m_{i,  i={1,2}} $  is  black hole mass  and $R_{\mathrm{in}}$  is  inner radius. 

 There are several empirical definitions of the radius of an accretion disk 
\citep{2002ApJ...573..754K},  and some more for slim  accretion disks \citep{2010A&A...521A..15A}. In our work   the inner radius is defined  as $R_{\mathrm{in}}\propto10R_{\mathrm{g}}$, where $R_{\mathrm{g}}$ is   the half of the Shwarzschild radius,   because it emphasizes the inner most place from which 
the UV/optical/IR luminosity originates.
Moreover, we also consider  that the inner radius corresponds to  the innermost
stable circular orbit (ISCO).

To estimate the outer radius $R_{\mathrm{out}}$   in units of light days, we 
adopt the relation given by \cite{Vicente14},  which is coming from the   microlensing observations of quasars:
\begin{equation}
R^i_{\mathrm{out}[ld]} = \frac{1}{2}\cdot r_0(\frac{m_i {\rm [M_{\odot}]}}{10^9})^{2/3}.
\label{eq:rho_out}
\end{equation}
where $r_{0}=4.5(\pm ^{0.7}_{1.6})$, 
and mass $m_i$ is given in solar masses.
 The outer radius of   the accretion
disks around black holes  in a compact binary system on a circular orbit,  could be  
tidally truncated  \citep[see][]{1980AcA....30..237P,1977MNRAS.181..441P,2014ApJ...785..115R}.  
We also consider this scenario,  setting  the outer radius of the  disk of the more massive component to
$\mathrm{R_{out-\mathbf{lc}}}\sim 0.27 q^{-0.3} a $ and of the  less massive component to $\mathrm{R_{out-\mathbf{sc}} }\sim 0.27 q^{0.3} a$,
where $a$ is a separation of components, $q=\mathbf{\frac{m_{1}}{m_{2}}}$ is the  mass ratio  of components,  and $\mathbf{m_{1}\leq m_{2}}$. The ratio of outer radii inferred from Eq. \ref{eq:rho_out}   ($\frac{\mathrm{R_{out-\mathbf{lc}}}}{ \mathrm{R_{out-\mathbf{sc}}}}\sim 
q^{0.67}$)  is almost the same  as   in the case of   tidally truncated  binary  system considered 
above ($\frac{\mathrm{R_{out-\mathbf{lc}}}}{ \mathrm{R_{out-\mathbf{sc}}}}\sim q^{0.6}$). Thus,  
Eq. \ref{eq:rho_out} can be adopted  for calculating  the disk dimensions.

 The emission from both  disk has a black body distribution and   the  summarized  luminosity $L(\lambda)$ at wavelength $\lambda$  from all parts of disk with different  effective temperature  is  given with:
\begin{equation}
L(\lambda)\propto \int_{S_{\mathrm{disc}}}\lambda dL(\lambda,T_{\mathrm{eff}})
\label{eq:SED}
\end{equation}
where $S_{\mathrm{disc}}$  is the area  of  the considered disk.

Due to the loss of energy, black holes  in a binary system approach  each other over time. Consequently,  mutual interaction between  one disk and opposite black hole component arises. This interaction can perturb  the disk temperature profile, causing the luminosity variation. 
Also, in compact  binary systems (where the distance  between black holes  is smaller than  0.1 pc),  the radial velocities of components can  increase to  the relativistic values. In that case, the effect of relativistic boosting can have an important influence. 
Both of those effects are taken into account in our model. Their detailed descripiton is given in   \cite{2015Natur.525..351D,Simic16}. Our model will be described in more details in \cite{Popovic18}. 

With this dynamical model we are able to reproduce light
curves for SMBHB systems with different parameters. As  an example we 
present in Figure  \ref{fig:allComponents1}  brightness variation for the object PG1302-102.
In this case we  take the time evolution of  the proposed binary system for four full orbits,  for which  we use a grid of 200 computational points, although
 a  higher number can be  considered. We test various models to roughly fit 
the observed
PG1302-102 light curve, and find that a  model with  the following 
parameters:  $m_{1} = 10^{8} M\odot$ ,
$m_{2} = 10^{9} M\odot$,  R = 0.015 pc, $\theta = 45^o$, $e = 0$, and orbital 
period $P =  1899$ days, can nearly describe the PG1302-102 light curve in the first  period. 
The light curves during four orbital periods of each component are 
shown in plot (a) and the resulting (total) light curve in plot (b) in  Figure  \ref{fig:allComponents1}.
 As one can see, there is a phase shift of local  extrema of  components, due to the opposing radial velocities. This indicates that  the relativistic boosting plays  a dominant role in such case, and that  the mutual interaction is almost negligible. We also see that  the variability of   one component is higher then the  total luminosity variation, especially in the case of  a less massive component.
The pure dynamical model, cannot explain the  part of PG1302-102 
light curve beyond 5000 days 
\citep[see][]{2018ApJ...859L..12L}, therefore we consider that an additional  attenuation in
the brightness of the SMBHB should be present.

\subsection{Perturbation in an accretion disk}\label{sec:gauss}

One  purpose of our model is to simulate 
 the long-term variability in SMBHBs. The variability can be caused by the 
 dynamical  parameters of the system, as it is shown in Figure  \ref{fig:allComponents1}, but   also by the
intrinsic variability   of one of the   components.  As 
often observed  in the  light curves of single   active galactic nuclei (AGN),  under  long-term monitoring,  the flux perturbations are present in the form of   outbursts  \citep{2010A&A...517A..42S, 2017MNRAS.470.4112G},   long-lasting flares as in the case of binary black hole candidates  NGC 4151  \citep[see Fig. 2 in][]{2008A&A...486...99S}  and
E1821+643 \citep{2017MNRAS.466.4759S},   or as remarkable low states  or minimum states characterized by an exceptionally weak continuum and line fluxes, also  seen in the case of binary black hole candidates NGC 4151  \citep{2008A&A...486...99S} and NGC 5548   \citep{2007ApJ...662..205B}.
The long term variability of some objects  has been successfully modeled by the variety of
disk perturbations ranging from the precession of  an elliptical disk,  or a disk with a spiral arm,  to bright spots, highly-ionized fast accretion  disk's outflows,  as well as   rotated, sheared, and decayed bright  spots \citep[see][and references therein]{2010ApJ...718..168J, 2012A&A...538A.107P}  and cold spots \citep{2017MNRAS.470.3027K}.

The  perturbations are well localized in the light
curve, which possibly reflect  the sharp edges of the emission region, 
 and are usually  resembling  to a Gaussian-like form. If  the angular dimension of  the emission region is much smaller than  the  viewing angle, a distant observer could not detect  the anisotropy of its radiation.
 Based on the above reasons, the  perturbations  in the light curves  have been modeled with Gaussian, exponential, and various other functional forms \citep[e.g.][]{1999ApJS..120...95V, 2011MNRAS.415.1631K,2015A&A...575A..55A}. 
\cite{ 2009JSMTE..02..051K} and \cite{2011ApJ...730...52K} proposed models of light curves based on superpositions of exponentially decaying  perturbations occurring at random times and with random amplitudes (the  latter  is also recognized as  Gaussian Process Regression). 
Gaussian-like perturbations could arise from  the intrinsic
quasar variability, for example  the  convergence  of  the  Poissonain process  once applied to the AGN light curve  as reported by \cite{2008A&A...487..815P,  2013A&A...556A..77P} can be understood in  the general statistical sense as  convergence to the Gaussian random process \citep{F17}.
Based on previous discussion, there is some evidence that   the Gaussian-shaped  perturbations are 
seen in AGN and even in magnetohydrodynamical simulations, though it is not the only type that can emerge.  
 
 Usually considered perturbation is in the form of an outburst, which  is obtained directly from  the magnetohydrodynamic simulation with a hot spot in the disk \citep{1991ApJ...376..214B, 2003MNRAS.341.1041A, 1992Natur.356...41A,  
1999MNRAS.306L..31P,2002MNRAS.333..800Z, 2004A&A...413..173N,2010MNRAS.402.1614D}, as well as with   a multicomponent spot settings     \citep{2006ApJ...651.1031S, 2008A&A...487..815P,  2013A&A...556A..77P,  2010A&A...510A...3Z}. 
 It is important to note that, as reported  by \cite{ 1999ApJS..120...95V},  the  magnetohydrodynamic models can  explain the physics underlying the flare  appearance and  its overall shape but   cannot provide its exact functional  time dependence   as phenomenological models  can.

  In the case of PG1302-102,  the drop in the brightness is seen, resembling the form of an inverted  Gaussian-like flare  \citep[see Figure 1 in][]{2018ApJ...859L..12L}, suggesting the local temporal decrease of the  disk temperature. This feature  in the light curve could be associated with cold spots \citep{2017MNRAS.470.3027K} or even with relatively small, dusty, rapidly-moving clouds partially covering the continuum and broad line region of a quasar  
\citep{2018MNRAS.478.1660G}.  Cold spots could be intuitively understood as  relatively confined subluminous regions like sunspots  \citep{1997ApJ...483L..13G}. The cold clumps could form naturally as a result of thermal instability in the hot gas \citep{2003ApJ...594L..99Y}  or condensation of the hot flow \citep{2000A&A...360.1170R, 2007ApJ...671..695L, 2007A&A...463....1M, 2007MNRAS.376..435M, 2009A&A...508..329M, 2011ApJ...726...10L}.

Thus, the Gaussian profile is taken in our model for simplicity,  and we assumed that  the        
   appearance of a cold spot causes the flux attenuation in the light curve of PG1302-102.
Moreover, if we consider asymmetrical perturbation, i.e Poissonian, it would affect the shape of  the modeled  inverted hump in the light curve, by producing an asymmetry on the  descending slope as  two stacked peaks. However, such feature is not supported by the observed data of PG1302-102.
The reason for this might  be found in differences between  gradients of Poissonian slopes and  those in the observed  inverted hump. Namely, a  Poissonian  descending slope  is much slower than the corresponding slope in the observed hump in PG1302-102. However, a Gaussian perturbation is symmetric with much faster descending slope, which is in better agreement with  the data, and supports  our  assumption to  use Gaussian perturbation.

\subsection{Gaussian-like disk perturbation - model} \label{sec:f}
 Following the previous section discussion, in our phenomenological model we include 
 the conceptual  Gaussian-like perturbation.
 To model  the  Gaussian-like perturbation in one component, we propose
 the temperature perturbation of the disk in more massive black hole  (as it  is shown 
in Figure~\ref{fig:allComponents2}, upper panels).
It is proposed that  the  whole body of the disk  is perturbed, where the
perturbation  reaches an  absolute  extremum  value  of approximately   $1.7\%$.
Applied perturbation  changes in time the disk temperature profile ($T_{\mathrm{{eff}}}^{\mathrm{pert}}(R,t)$) according to the following expression:
\begin{eqnarray}
T_{\mathrm{{eff}}}^{\mathrm{pert}}(R,t)=T_{\mathrm{eff}}(R) + T_{\mathrm{eff}}(R)\cdot\delta T(t),  \\
\delta T(t)= P_{\mathrm{int}}\cdot\exp\left[-\frac{(t-t_{\mathrm{pert}})^2}{P_{\mathrm{dur}}^2}\right] 
\label{eq:T_pert}
\end{eqnarray}
\noindent where $T_{\mathrm{eff}}(R)$   is  the  unperturbed disk temperature profile.
 Note that  a multiplication in the time domain is equivalent to  a convolution in the frequency domain. The Gaussian kernel is the physical equivalent of the mathematical point. It is not strictly local, like the mathematical point, but semi-local. Over its  lifetime, the  perturbation produces a coherent temperature perturbation sampled by a window function $\delta T(t)$.
  The sign of   the  intensity of the temperature  perturbation, $P_{\mathrm{int}}$, determines whether   it  is a magnification (positive sign) or  an attenuation (negative sign).
Our numerical tests confirmed that   an inverted Gaussian-like temperature perturbation  results in the  inverted Gaussian-like shape of  luminosity curve. 
 The perturbation is applied on the disk temperature profile of the more massive
component and than superposed with  an emission from  the less massive black hole.
As we can see in Eq. \ref{eq:T_pert}
the perturbation  decreases temperature at instant $t_{\mathrm{pert}}$,
for amount $P_{\mathrm{int}}$ and  duration $P_{\mathrm{dur}}$ (  Figure~\ref{fig:allComponents2}).

 The parameters of  the perturbation are found
  when comparing  the modeled and  observed data,   using  the condition of
minimization of statistical parameters  which defines  the goodness of the  fit.
 Here we intentionally model  the Gaussian-like
perturbation
using three free parameters (instead of two) in order to allow more
flexibility in  the fitting procedure. 

\begin{figure}
\centering
\includegraphics[width=0.6\textwidth]{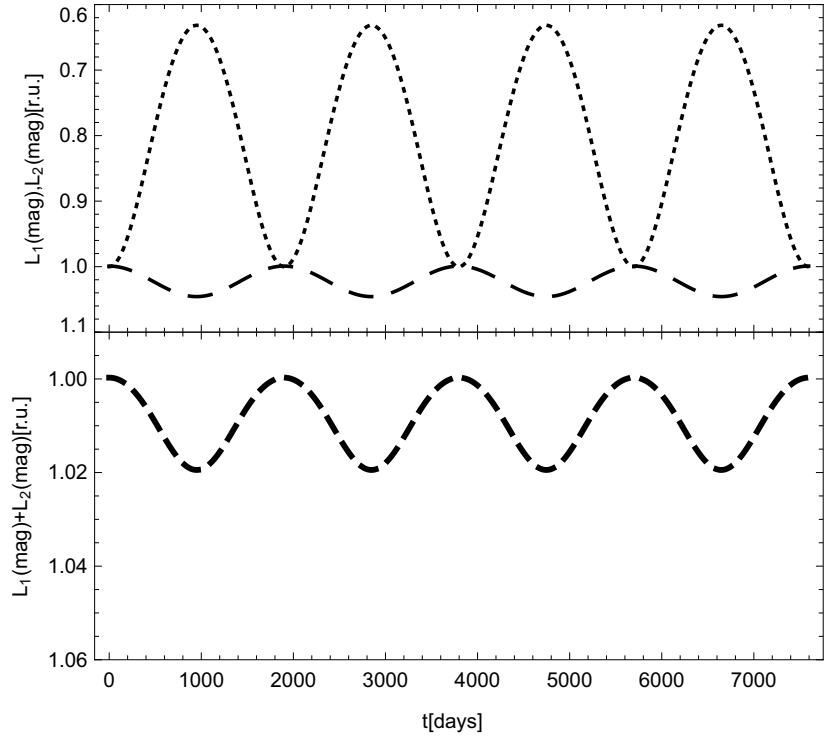}
\caption{
Modeled light curve of the SMBHB system during four orbital periods (see 
text): (a) Individual light curves  (L1, L2) of
 the corresponding  accretion disks of components $\mathrm{m}_1$ (doted) and $\mathrm{m}_2$ (dashed);  (b) The
modeled light curve of the total luminosity ($\mathrm{L}1 + \mathrm{L}2$) emitted from the 
SMBHB. The luminosity   is given in relative units on y-axis, and time is given 
on x-axis  in days.}
\label{fig:allComponents1}
\end{figure}

\begin{figure}
\centering
\includegraphics[width=0.6\textwidth]{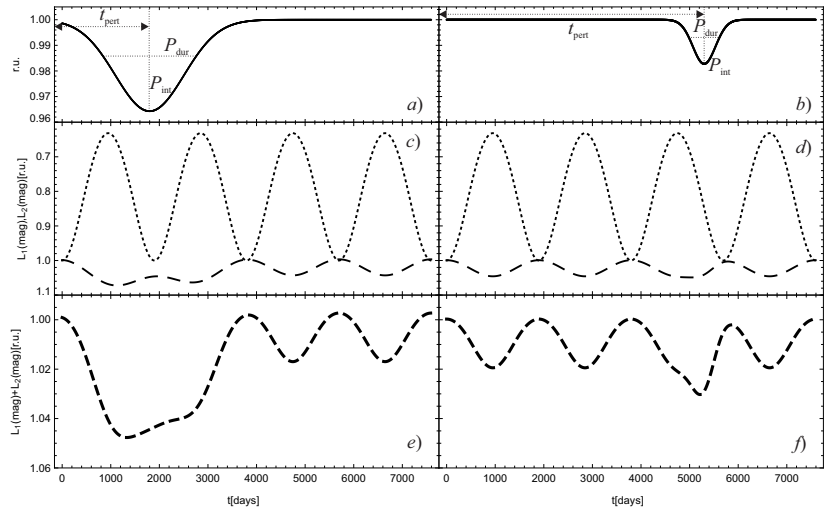}
\caption{
The influence of  the different perturbations (the shapes are present  in upper panels (a) 
and (b), see text)
on the light curve of more massive  component (see  middle panels  (c) and (d) ) and the 
resulting light curves (shown  in  bottom panels (e) and (f) ).}
\label{fig:allComponents2}
\end{figure}

In Figure~\ref{fig:allComponents2},  for example two hypothetical perturbations are present: 
first, that occurs 1800 days 
after the beginning  of  the monitoring,  with  an extremum  intensity  of $P_{int} =
3.5\% $  of the
total disk emission, and duration  of 1000 days (upper left panel  (a)); the influence on 
the
more massive component light curve  (middle left panel (c)) and the SMBHB light curve (bottom left panel (e)) are also presented. Also, we explore  a more realistic 
perturbation
at around   5300 days after the first observation, lasting for   330
days and with lower
intensity $\mathrm{P}_{\mathrm{int}} = 1.7\%$. Its  shape  and the 
effects are given  in Figure \ref{fig:allComponents2} (right panels  (b) and (d)). As can be seen from the bottom panels, a 
perturbation in the
disk of one of the SMBHB component can significantly deform  the periodical 
shape of  the total SMBHB
light curve (compare  (b) panel in Figure \ref{fig:allComponents1} with the  bottom panels (e) and (f) in
Figure \ref{fig:allComponents2}).

Summarizing, our SMBHB model provides  the following parameters:
black hole masses  $m_1$, $m_2$,   $m_{1}\leq m_{2}$, their  separation $a$, inclination of  their orbital plane $\theta$, eccentricity $ e$  which is the same for   both black hole orbits, orbital period $P$, time  $t_{\mathrm{pert}}$ when  an extremum  occurs in the disk of $m_2$, intensity $P_{\mathrm{int}}$, and  the duration at half of  the perturbation $P_{\mathrm{dur}}$.

\section{Data and Method} \label{sec:datmet}

\begin{figure}
\plotone{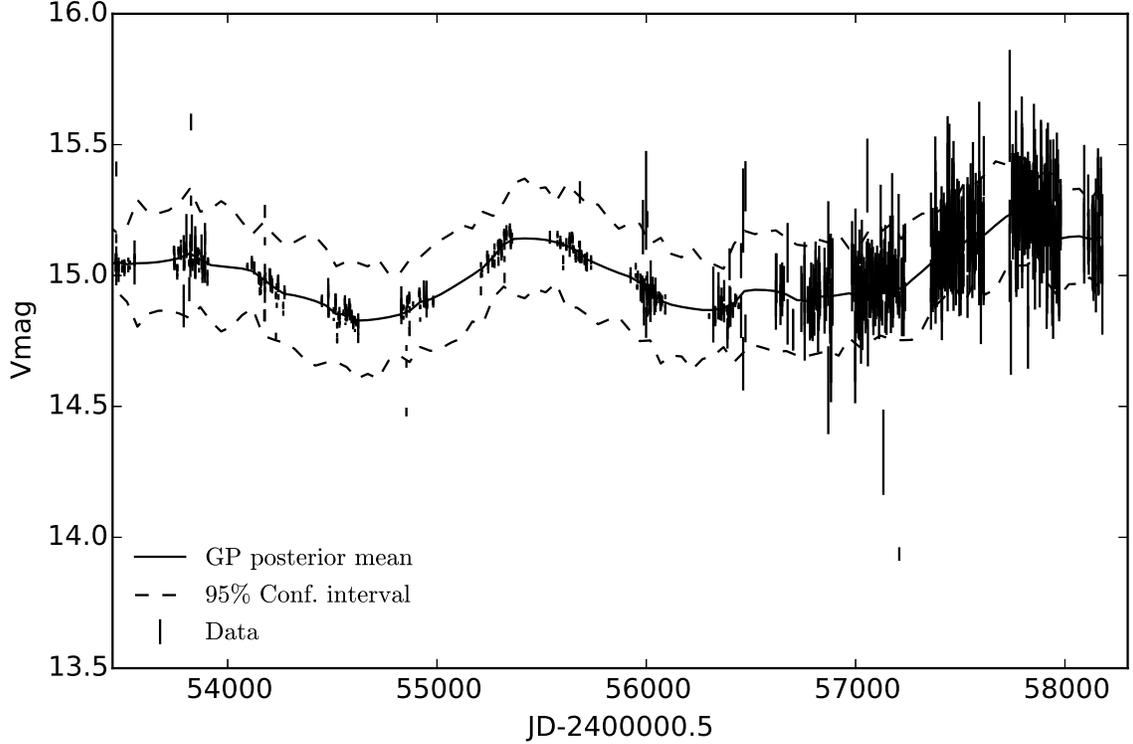}
\caption{The best fit with a nonstationary GP mean as a solid line with $95 \%$ of Confidence Interval between dashed lines.  The observed photometric obseravtions  are given as vertical error bars.\label{fig:curve22} }
\end{figure}

 In this study we use  the photometric light curve of PG1302-102 collected by LINEAR, CRTS and ASAS-SN surveys  that were employed for periodic analysis
  reported in \citet{2018ApJ...859L..12L}. A detailed description of the data sets can be found in  \citet[][and references therein]{2018ApJ...859L..12L} and will not be repeated here. 
  In order to  apply our hybrid method for periodicity detection, we  mitigate possible  effects of the gaps within the light curve,  and we  thus pre-process the photometric  light curve by modeling with a robust Gaussian Process Regression (GP, machine learning) method as reported in   \citet{ 2018MNRAS.475.2051K}. 
Here we use a GP with a non-stationary kernel  to fit data, which is obtained by  the standard procedure of summation  of quasi-periodic and Ornstein-Uhlenbeck (OU) kernels \citep{2017Ap&SS.362...31K}. The modeled GP light curve is given in Figure~\ref{fig:curve22}. 

We use  the  hybrid method  reported  in \cite{2018MNRAS.475.2051K}  to determine  the periodicity in PG1302-102 time series.
The hybrid method, thanks to combination of  two well-developed techniques in common use, continous wavelet transform ( CWT) and correlation coefficients,  is an easily applicable procedure to the problem of periodicity. Its  key advantage over other techniques is that it does not require   any assumptions about the stationarity of the data.

  The 2D  correlation map of periodicities  can be calculated in two ways, either using 
identical or different data sets  for an input, as it is  a practice in generally similar technique of 
2D correlation spectroscopy \citep{Noda15}. 
 The 2D correlation map of identical data  sets deconvolves and determines correlations between periodicities 
 in one light curve.  In the case of PG1302-102 we use this approach due to availability 
 of a single light curve.  The 2D correlation map involves evaluation of the envelope of CWT of the light curve, after which  a nonparametric Spearman's rank correlation
 for all possible pairs of the values of the envelope  is calculated, generating two-dimensional matrix 
 of negative and positive correlations. 
 Note that due to  the  normalization of correlation coefficients,  
  the correlation coefficient intensity at the main diagonal position is of  the order 1, 
   thus  influencing  the prominence of correlation clusters which indicates presence of oscillatory patterns. 
    The diagonal feature in the case of identical data sets is more evident,
     and  in the case of different data sets with specific  physical dynamics and/or observational characteristics, 
      diagonal can be broken and/or correlation clusters detached. Moreover, 
       in the case of perfect correlation over all periods, the topology of the map would resemble a homogeneous cone, 
       with an  apex and an open-end in the lower left and  the upper right corner, respectively. 
 Generally, as in the 2D correlation spectroscopy, the noise in the data affect correlation clusters to appear  broader and smeared.

After applying  the hybrid method on observed light curve, we perform consistency check of
  the     result  supported by non linear fitting sinusoid, which  has the form of 
\begin{equation}
V_{\mathrm{mag}}=A\sin (\frac{2\pi t}{P}+\varphi)+B,
\label{eq:sinusoid}
\end{equation}
where $A$ is amplitude,  $P$ is period, $\varphi$ is phase  and $B$ is offset. 
The  best fit was derived based on the reduced $\chi^2$. 

  Since one of the goals of our analysis is to model  the  perturbation of the periodic signal in the light curve of PG1302-102, it is important to distinct the meaningful signal from the noise, which color is  not  known {\it a priori}. A specific test for that noise color must be applied, which we describe in more details and apply on both observed  and GP modeled light curve in the following subsection. 
   
\subsection{Noise test for the light curve}\label{sec:nnoise}

  A  red noise process can be interpreted as an  autoregressive process of the first order AR(1),  with positive correlation at unit lag.
 A  pink noise can be modeled   the the differencing parameter d($=0.5$) of the Box -- Jenkins autoregressive integrated moving average (ARIMA) strategy, taking on it continuous values \citep{Box70}.
Autoregressive fractionally integrated moving average (ARFIMA) modeling improves the  Box–-Jenkins approach by implementing  the differencing parameter d to have  non-integer values. This allows ARFIMA to fit any long range memory in time series remarkably \citep{Brock02}.  
However, these processes are  stationary.
If the PG1302-102 light curve is found to be non stationary, then it is different from mentioned noise processes. If the light curve is stationary further statistical procedures must be applied to verify that the light curve is  a noise process.
Thus, a stationarity test is first applied  on the given light curve.
For this purpose we used Kwiatkowski–-Phillips–-Schmidt–-Shin (KPSS)
test \citep{DOI: 10.1016/0304-4076(92)90104-Y}. The test can be applied on time series with gaps, because simply ignoring gaps or filling them with interpolated values does not  alter asymptotic results associated with its statistics.
In our analysis the KPSS test, applied to the  observed light curve with ignored gaps and  to the  GP modeled curve, rejects stationary null hypothesis in favor of the non-stationarity alternative at $5\%$ significance level.  
As the result of the test,   the signal is discerned  as a subsequences of light curve differing from the noise.
Furthermore, we  add   the white noise to our modeled data  in Section \ref{sec:modelsmbh}, since the autocorrelation functions of  each cluster of points in  the observed light curve correspond to the  white noise. It  is generated  as a  random process with a Gaussian distribution, with the mean value  of  $\mu=0$ and the absolute value of width $\sigma$, which  depends on units used. In our computation we normalize flux to 1, so that parameter $\sigma$ has value around $0.5\%$, i. e.  $\sigma=0.005$. 

In order to investigate  the effect of perturbations and added white noise to the modeled  light curve, we perform  the following numerical experiment.  We apply our hybrid method on the hypothetical curve with and without  the white noise
 (see  bottom right plot  in Figure  \ref{fig:allComponents2}). Our hybrid method gives the period  of $1873 \pm \ 250$ days in both cases, when white noise is included and excluded. This  is in  agreement with the period of  1899 days inferred from the model (Section \ref{sec:modelsmbh}). In both cases we could detect the presence of periodicities. However, the uncertainty is large due to  large  magnitude of perturbation.

\section{Results and Discussion} \label{sec:resdisc}

\begin{figure}
\centering
\includegraphics[width=0.8\textwidth]{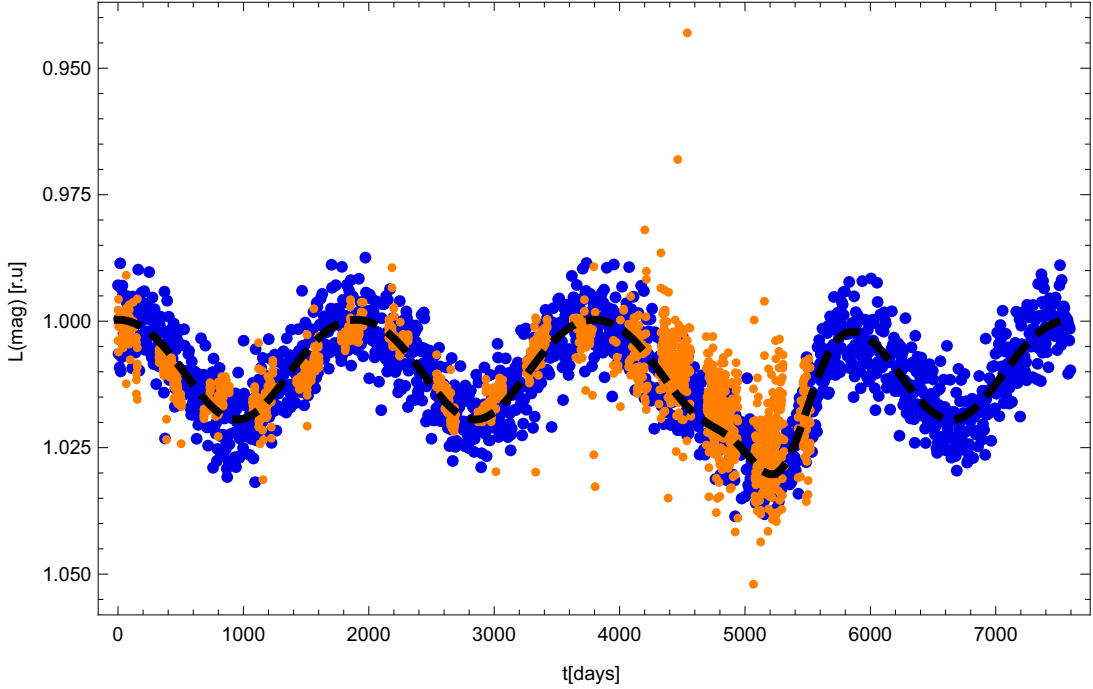}
\caption{Observed  (orange points) and modeled light curve (blue points). Dashed black line represents  the modeled curve without white noise. Time is  given on x-axis in days, whereas the flux ( note that the
observed light curve is previously expressed in magnitudes) in relative units is given on y-axis.}
\label{fig:dataModel}
\end{figure}

Using  the proposed model  from Section \ref{sec:modelsmbh} we first inferred  the observed  light curve without white noise (see right plots  in Figure \ref{fig:allComponents2}).  Then, the modeled light curve  given in  Figure \ref{fig:dataModel} is obtained by adding the white noise to the model.    As we can see, there is a very  good agreement between the observations and model with perturbation for the  set of  inferred parameters given in Table \ref{tab:parameters}.

\begin{table}
\begin{center}
\caption{ Inferred parameters  of the model of the SMBHB system with Gaussian perturbation in the accretion disk of the  more massive component,  defined  in  Section \ref{sec:modelsmbh}.
 Parameters AIC,  BIC,  $\mathrm{AIC_{np}},\mathrm{ BIC_{np}} $, and $\mathrm{AIC_{nc}},\mathrm{ BIC_{nc}} $ measure the quality  of perturbed, non-perturbed   and pure noise model, respectively (see text).} \label{tab:parameters}
 \setlength{\tabcolsep}{4pt}
\begin{tabular} {|c|c|c|c|c|c|c|c|c|c|c|c|c|c|}
\hline
$m_1$&$m_2$ & $a$ &  $e$   & $t_{\mathrm{pert}}$& $P_{\mathrm{int}}$& $P_{\mathrm{dur}}$& $P$&AIC&BIC& $\mathrm{AIC_{np}}$&$\mathrm{ BIC_{np}}$& $\mathrm{AIC_{nc}}$&$\mathrm{ BIC_{nc}}$ \\
 $[10^{8} M_{\odot}]$& $[10^{8} M_{\odot}]$&[pc]& &[days]&[$\%$]&[days]&[days] & & & && &\\
\hline
 1 & 10&0.015&0& 5300& 1.7&330&1899&-4135&  -4125& -3793& -3787 & -3028& -3025\\

\hline
\end{tabular}
\end{center}
\end{table}

The model with perturbation produces orbital period of $P= 1899$ days,  and 
circular orbits. Inferred larger mass and mass ratio is consistent  with those obtained in analysis of \cite{2015Natur.525..351D}, while the period estimate  is slightly  larger for about  10 days then reported in \cite{2015Natur.518...74G}, but still within the error bars.  Setting inner radii of both disks to the value of ISCO does not affect the simulated light curve, because 
the regions  close to the black holes  radiate photons of  much higher energies which do not contribute to the   observed  band directly. The assumption that the outer radii of the  larger  and the  smaller disks are defined by the tidal truncation (see Section \ref{sec:modelsmbh}),  produces negligible variation in the light curve amplitude  of approximately several percents. 
 
  We used Eq. \ref{eq:rho_out}  to calculate the outer radius of both accretion disks, 
which  ratio is
around 4,  that is very close to the value of 4.6 in the case
of tidally truncated binary system,  taking into account $q=0.1$. 
The difference between these two ratios  reflects  difference  in the 
luminosity of several percents, additionally, we considered only the 
perturbation in the  more massive component.

We compute the AIC (Akaike Information Criterion) and BIC (Bayesian
Information Criterion) parameters which define criteria for model selection
and effectiveness   (Table \ref{tab:parameters}).  In  the computation,  our model produce  simulated  light curve for the  same time points as given in the observed light curve.
Also, in order to have more realistic values of AIC and BIC parameters,
model points are computed at the same moment when data points are recorded.
Small variations for AIC and BIC values are possible since stochastic
nature of included white noise. 
 Note that  AIC and BIC of our models differ significantly from those
 obtained by  \cite{2018ApJ...859L..12L}. 
 This  could be due to different normalization of
  data as well as  due to different models. \cite{2018ApJ...859L..12L}  model is a
 rational function of short-timescale variance  and characteristic time scale (see their equation 4). 
 We calculated the  log-likelihood function (L)  of the data $y_{n}$ given the parameters $\theta$ as follows

\begin{equation}
\mathrm{ln} L = -\frac{1}{2} r^{T} K^{-1} r-\frac{1}{2} \mathrm{ ln} |K| -\frac{N}{2} \mathrm{ln} 2\pi
\label{eq:aic}
\end{equation}

\noindent where $r$ is the residual  vector between the mean flux predicted in a model and the observed flux at 
each observation time $t_{i}$. Here $K={\sigma^2_{ij}} \delta_{ij}$, $\delta_{ij}=1, i=j $ and $\delta_{ij}=0, i\neq j$, is the diagonal covariance matrix where {\boldmath$\sigma^2_{ij}$}   is
the variance. The fact that $K$ is diagonal is the result of our 
earlier evidence that the  the data are uncorrelated.   The variances are obtained as  second derivatives of the likelihood function with respect to the model parameters  fitted  to the data (i.e. as diagonal elements of the Fisher information matrix). 
Note that due to different   modeling approach,
\cite{2018ApJ...859L..12L} applied their  method only to the binned light curve, which consisted of  19 and 35  barricentric points 
for LINEAR$+$CRTS  and LINEAR$+$CRTS$+$ASASSN data set respectively, assuming that these  barycentric  points are correlated
as damped random walk. The same assumption of  correlation between data points was used in \cite{2015Natur.525..351D}
but the data set consisted of about 250 points. 
This implies that the covariance  matrices $K={\sigma_{i}} \delta_{ij} +k(t_{i},t_{j})$ had  non diagonal elements
 $k(t_{i},t_{j})$ arising from
 the assumed damped random walk in these studies. 
 
  We note that \cite{ 2015Natur.525..351D} and \cite{2018ApJ...859L..12L} analysis  did not use Fisher Information matrix as an estimate for their covariance matrices. Instead, former study used  log likelihood covariance matrix with added variances  in the photometric measurement  on the diagonal,  and Liu et al (2018)  had the Gaussian noise added on the diagonal   which could explain AIC differences about 10  in their calculations.

Thus, the major difference between our AIC, BIC  and those found in the recent papers most likely 
lies in the form of covariance matrix constituting the Gaussian likelihood function as well as in the number of used data which 
 affect the dimension of the matrix and later the process of   maximization.
Our approach is distinct from previous  ones in the fact that complete data set of 1700 points which have white noise characteristic  is modeled without binning.

 Besides general differences in covariance matrix,  note that the data of \cite{ 2015Natur.525..351D} does not show attenuation, and that in \cite{2018ApJ...859L..12L} study perturbation was not included in their model
and due to the binning of the light curve the hump in the light curve was  covered by  15  barycentric points which slightly  changed the geometry of the hump. 
  The number of points
in the hump is about 873 which is almost  half of all available data implying  that large portion of information lies in the hump too.
Our model  shows a large difference between the pure noise and  perturbed model  confirming  the importance of
information confined in the hump of PG1302-102 light curve.

Moreover,  the differences  between information  criteria of  the composite  sinusoid-noise  and the pure noise model were 
close to  10  in previous
studies.   A model with AIC difference strictly  larger than 10 units of the best model,  which indicates that the model is less favorable, will have no support, and can be omitted from further consideration \citep{Burn02}. However,  models with differences up to 10  are  usually considered  as no superior to some other models in a set of considered  models. In this case model averaging (combining) gives a relatively more stabilized inference \citep{Burn02}. From this point of view  the previous analysis suggests a model combination.

We can see that our model as a composite, consists of  three parts:  dynamical (sinusoidal), perturbation, and noise part, and demonstrates
 that light curve is best described with combined model due to large AIC (BIC)  difference with respect to the pure noise model, which is 
 in agreement with findings of previous studies. Thus, for the first time, we confirmed that it is possible to model a complete data set of PG1302-102 without any binning, and extract a  valuable information of 
 a periodic signal and perturbation. 
 
As for the absolute magnitude of our AIC and BIC, we will note that in the maximum Gaussian log-likelihood solution
 the weights vector is identical to the least squares solution,
  which means that the better the independent variables of the model
   are in predicting the dependent variable, the more negative the AIC becomes.
    In ideal case it  would approach negative infinity. 
    As for the comparison of models, in the case of two models (as it is in our analysis) 
    the statistical rule is to compare only their AIC values (not  absolute value), thus model with lower 
    AIC  would be the preferred one.
 Here both AIC and BIC are the smallest  for the model with perturbation included.    
     In the case of more than two models  the
    difference between i-th model AIC and minimum AIC among all models is used. It is not possible to use this set of models approach to  compare our AIC with those  in \cite{2018ApJ...859L..12L}  analysis 
    due to different Gaussian-likelihood formulation and  possible  
    different  normalization of data.

Next, we apply our hybrid method  to the
modeled (Figure~\ref{fig:dataModel}) and preprocessed observed light curve (Figure~\ref{fig:curve22}), for which we tested that it is not  the  noise process (subsection \ref{sec:nnoise}).

\begin{figure}
\centering
\includegraphics[width=0.7\textwidth]{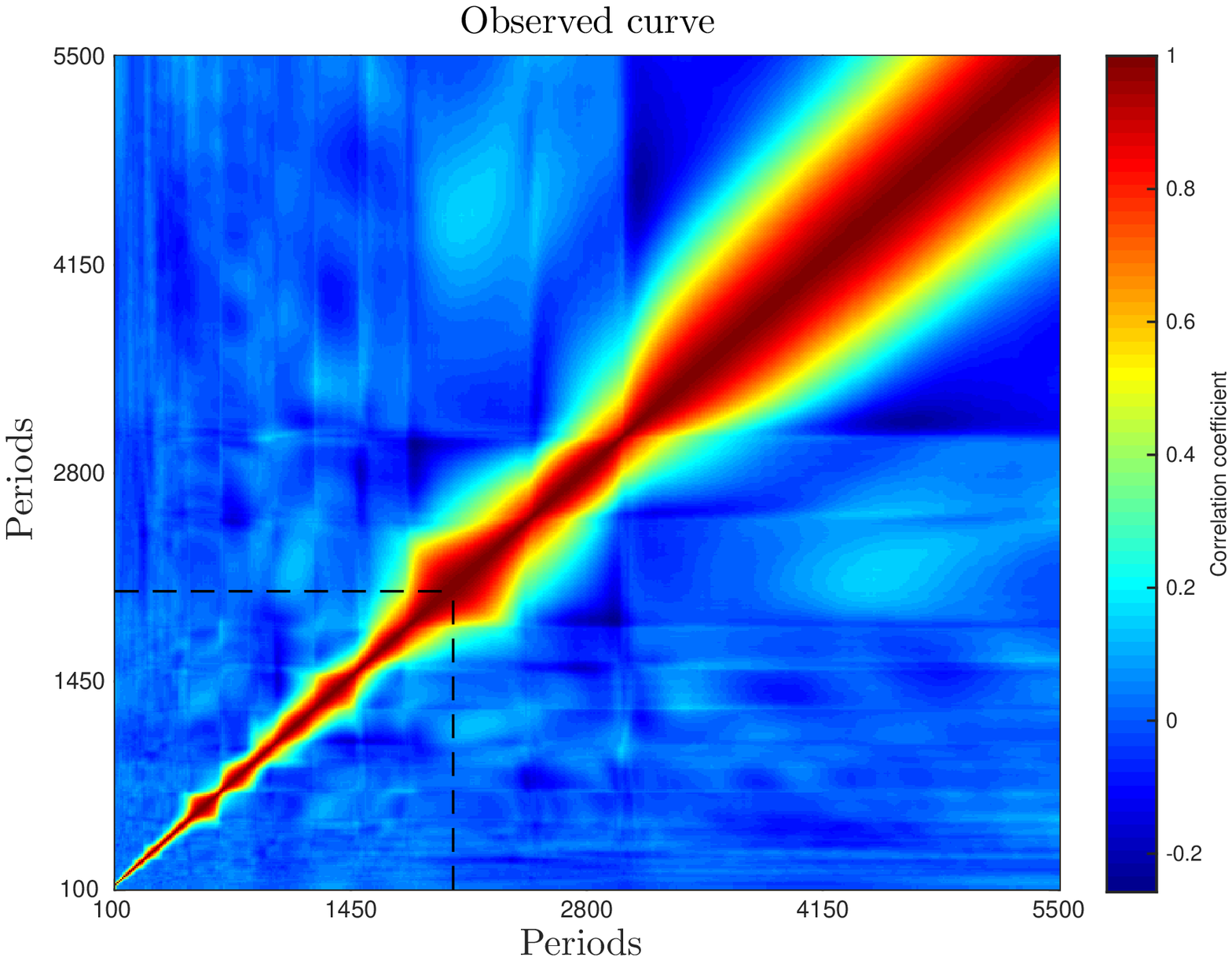} \\
\includegraphics[width=0.7\textwidth]{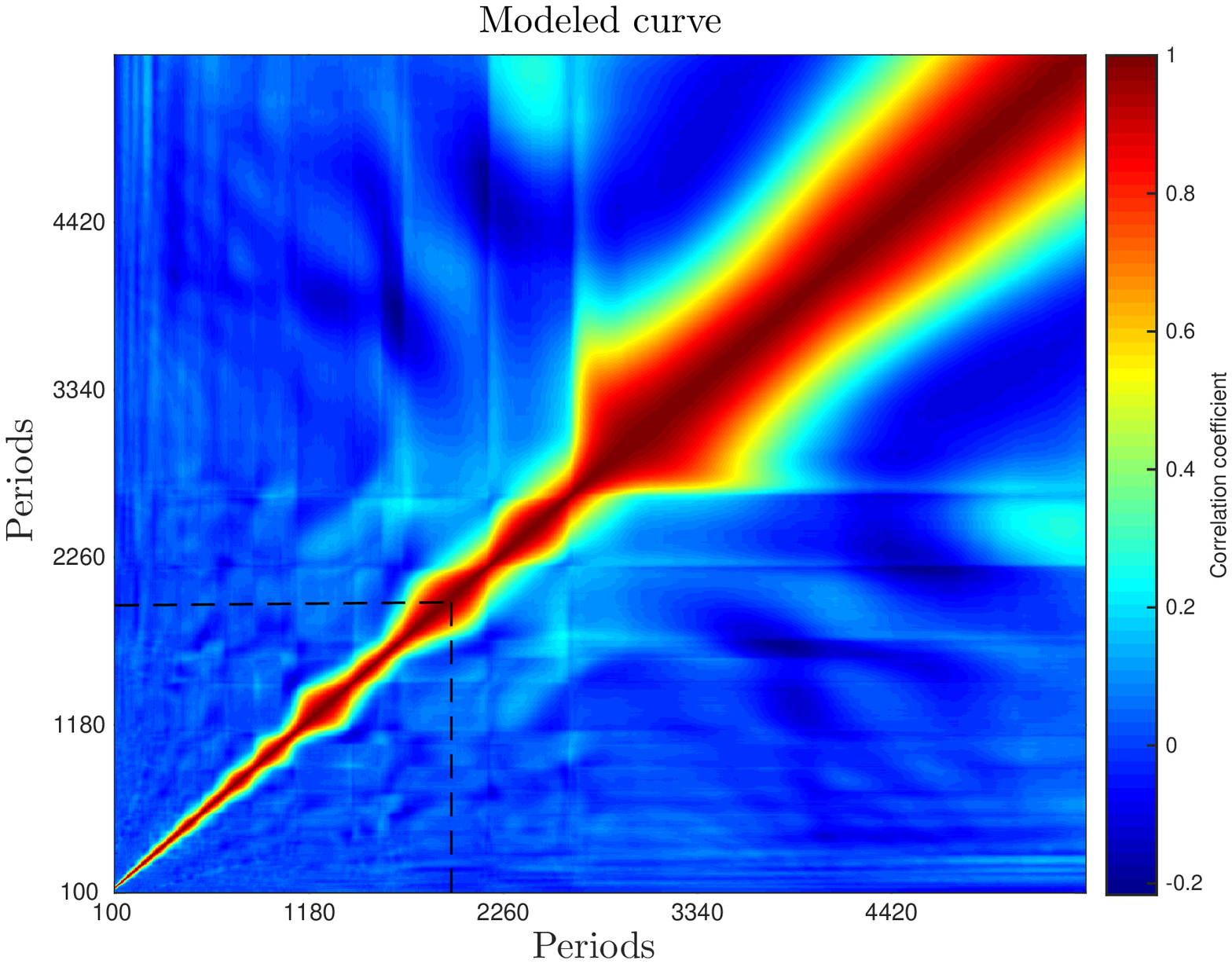}
\caption{   The 2D correlation map of  all oscillatory patterns within the total observing time  range $100-5500$ days, for
preprocessed  observed light  curve (top) and modeled  light curve (bottom).   Both axes represent periods (in  days) in the curve.  
Diagonal correlation clusters means that oscillations are caused by physical processes within PG1302-102. Values of correlation coefficients are given on colorbar. The clusters of high correlation are marked in  red with significance $p<0.00001$.
Dashed line marks detected period.
\label{fig:periods}}
\end{figure}

As can be seen in   Figure~\ref{fig:periods}, both curves have almost similar 2D correlation maps. 
  The topology of our 2D correlation map  is fragmented  and   attached correlation clusters are  visible.
There are  two  important regions of periods, the largest  about 4000 days on both maps and smaller clusters at   $1873 \pm  250$ and  $1972 \pm 254$ days,   for modeled and preprocessed observed curve, respectively.   Their correlation coefficients are 0.99 with a significance of $p<0.00001$.
The period of $\sim 4000$ days can be neglected since it corresponds to the whole observing period.
It is clear  that our method   has recovered the period of the modeled light curve successfully.
  The noise  is  present  especially in the hump of PG1302-102 light curve, which effect is evident on the appearance of the correlation cluster associated with this period.

 Also,  the hybrid method indicates that the periodicity of the observed light curve is very close to the one found in  the modeled one.
Interestingly, our hybrid method gives almost identical period for the modeled light curve without  white noise.

 As a further verification we fit  a sinusoid to the detrended (mean value is subtracted) observed light curve, which was selected based on reduced $\chi^2$ by being closest to 1.
    The obtained best fitting  parameters with their standard deviations are  given in Table \ref{tab:paramsinus}.

\begin{table}
\begin{center}
\caption{Best fitting  parameters and their standard deviations  of the sinusoid fitted to the detrended observed data,  defined by Eq. \ref{eq:sinusoid} in  Section \ref{sec:datmet}. The parameters $\chi_{\mathrm{red}}^2$ is reduced $\chi^2$  value of the fit.} \label{tab:paramsinus}
\begin{tabular}{|c|c|c|c|c|}
\hline
$\mathrm{A}$& $\mathrm{P}$ & $\varphi$&$\mathrm{B}$& $\chi_\mathrm{{red}}^2$\\
                     &[days]&[radian]& & \\
\hline                     
0.123 $\pm$0.003 & 1950$\pm$150 & 4.74$\pm$0.84&  -0.018$\pm$0.003&  1.09  \\
\hline
\end{tabular}
\end{center}
\end{table}

\begin{figure}
\plotone{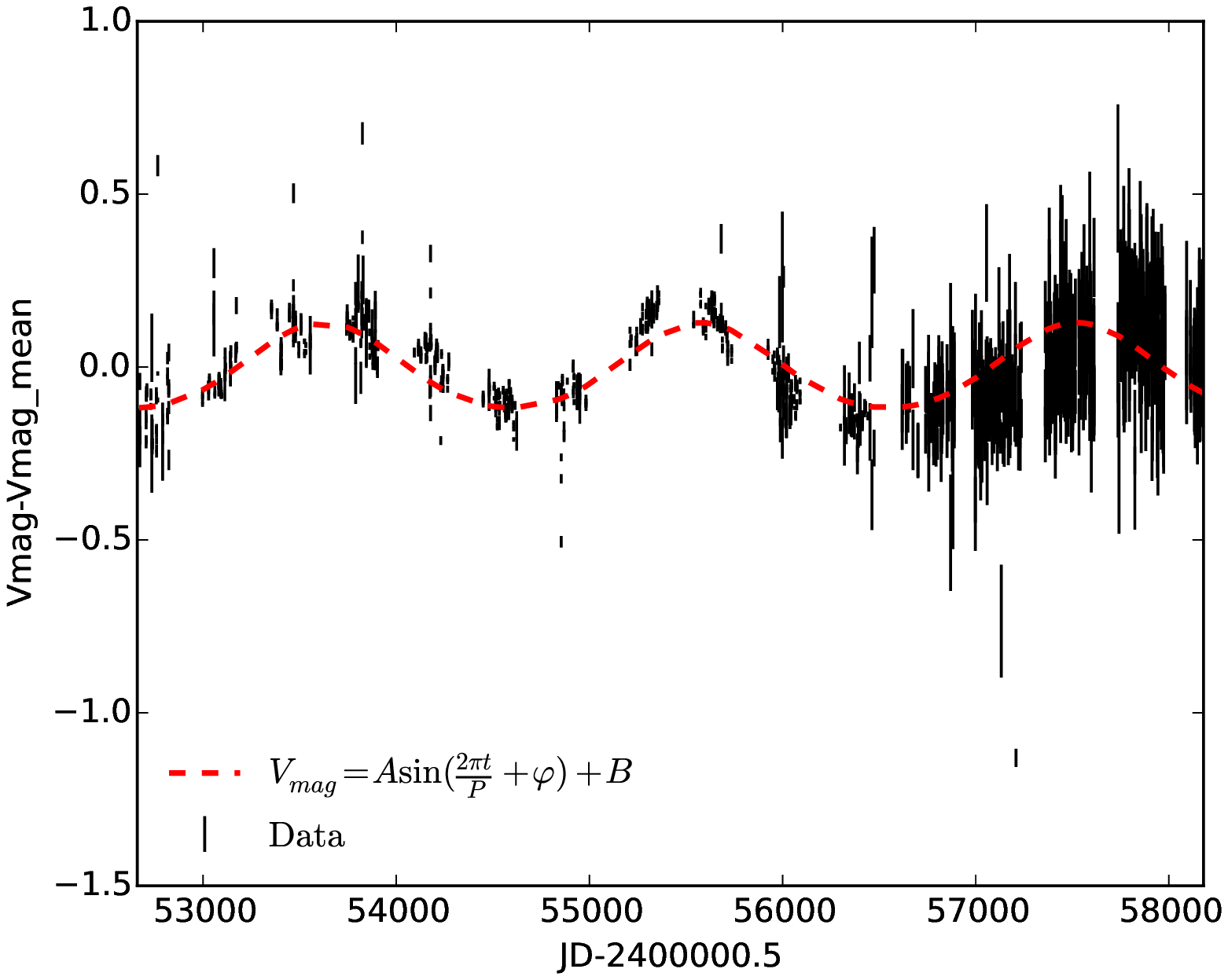}
\caption{  Best fitting of sinusoid  model to the detrended (mean value is subtracted) observed light  curve. Photometric magnitudes  are represented by  error bars whereas model with dashed, red line. The best fitting   parameters are  given in  Table \ref{tab:paramsinus}. \label{fig:sinefit}}
\end{figure}

Figure~\ref{fig:sinefit} shows sinusoidal model (see  Eq. \ref{eq:sinusoid} and  Table \ref{tab:paramsinus} ) nicely describing the peaks and troughs of the original data, with reduced $\chi^2$  value of 1.09.

The SMBHB model applied to the optical curve of PG1302-102 explains the variability of the optical flux and  its  perturbation leading to  slight changes in detected  orbital period of the system. The model recreates physical conditions in  the accretion disk of a  more massive black hole causing  attenuation and the dynamic properties of the system.  On the other hand, the model cannot {\it apriori} predict  the repetition of the perturbation, but we can provide some general statistical estimates following  the prescription given in \cite{2005ApJ...621..940S}. If we assume that  the characteristic  lifetime of the perturbation  is about  400 days, as it is  the  duration of  the inverted Gaussian-like perturbation inferred from our  model,   then the number of such events which could be expected  with  the  life time  between 400 and 450 days over the next 2000 days is about 1.7. Based on this pure statistical view,  there is a chance to detect  a similar event within next few years.

In our  analysis  we consider one time event, because  the  data did not show any similar feature up to now.
The recurrence of perturbations is well analyzed by
   the magnetohydrodynamic simulations directly producing the hot spots in the
accretion flow and implying  that these phenomena  could be either periodic (and destroyed 
 by differential rotation) or aperiodic  \citep{2003PASJ...55.1121F}.
Based on this, something similar could be expected for cold spots. Thus, if the  perturbation in PG1302-102 light curve  is periodic, there is also a possibility that  its   next amplitude would appear smaller  due to differential rotation which can gradually destroy it.

Moreover, \cite{2000ApJ...531..744V} pointed  to  a  possibility that at certain intervals
 the   less massive black hole crosses the accretion disk  of the  more massive black hole,  and  the effect of shadowing the certain parcel of the disk could be expected \citep{2013LRR....16....1A}.
  This means that the peak from  such event should be broader in time, 
  because the passing takes longer than the crossing. 
    Such dynamical relation (and its  signal) is likely to be periodic, but with periods  of the order of decades to centuries. Thus, we are likely to detect  separate events and anticipate them as isolated flares \citep{2008ApJS..174..455B}. 
     Moreover,  linking the circumbinary disk to the mini-disks  and  the gas density at the inner edge of the circumbinary disk, a lump has been
observed  in  the recent magnetohydrodynamic  simulations of binary black hole systems
\citep{2008ApJ...672...83M, 2012ApJ...749..118S, 2013MNRAS.436.2997D,  2014ApJ...783..134F,2015ApJ...807..131S, 2016MNRAS.459.2379D, 2017MNRAS.469.4258T, 2018MNRAS.476.2249T}.
  A generic result of  these simulations  is  a  low-density cavity created by the binary torque, but gas  can still leak into the cavity through non-axisymmetric streams,  implying lower temperature in  such region.   Recently, \cite{2013MNRAS.434.2275T} proposed that periodic streams into the cavity may activate  more noticeable variability than previously thought.

 However, the first  general relativistic magnetohydrodynamic  simulation \citep{2018ApJ...853L..17B}, assuming that the binary separation is relativistic,  revealed even more dramatic picture. Namely,  the response of
the  accretion  disks around black holes to the circumbinary disk  in the binary system may introduce distinctive time-dependent features in the binary's electromagnetic emission.

  Instead of a single perturbation,  multiple perturbations  can appear  in the disk, 
which may lead to  the superposition of their individual  luminosities into a complex signal, 
and a periodicity analysis is required. However, the structure of the data in the  inverted hump of
 PG1302-102 is not favorable of this scenario, and  the periodicity analysis could not disentangle  superimposed 
  signals if any present. 

  Perturbations  were recorded in the light curves of some well studied objects
as we mentioned in Subsection \ref{sec:gauss}. 
The most comprehensive study  to date,  specifically focused on  a systematic search for major flares in AGN, is by \cite{2017MNRAS.470.4112G}. They presented remarkable results of the CRTS,  identifying 51 events from the sample of  more than 900000 quasars, typically lasting 900 days.  The inverted hump of PG1302-102 is within this time range,  however the physical explanations suggested by \cite{2017MNRAS.470.4112G} such as single point single lens model, supernova or tidal
distribution events, are not applicable in  the  case of observed PG1302-102 inverted hump.

As we discussed, there is  some chance that  the cold spot event occurs in the future, which could affect the light curve over certain range of time. This event  could be superimposed on main sinusoidal signal of PG1302-102 binary candidate, but this main periodic signal could be still extracted  as it is shown by our hybrid method.

As we have already mentioned, there are  physical  possibilities for flaring appearances ranging from density irregularities within the disk,  up to dynamical reasons as it is a companion black hole producing shadow at   the disk after entering the disk. 

Moreover, companion black hole radiation could be also important \citep{1998ApJ...508..669P}.
This object could  experience increased accretion when it crosses the accretion disk, passes the pericenter, and crosses in front of the jet. All these instances cause increased flux via temporary accretion disk and jet formation in the companion.
There  is also another scenario related to circumbinary disk  (which may be a condition  for some binaries to overcome the {\it final parsec} barrier), having a large cavity \citep[see][and references therein] {2012EPJWC..3906008T}.
After  passing  through pericenter near one of the SMBHs, the stream  could  self-intersect and  produce a shock. This material would circularize into a hot, optically thick annulus and viscously spread. The gas will begin to accrete in a slim-disk form  \citep{2009MNRAS.400.2070S} before it can cool radiatively, producing flares in optical, UV and X-domain  at the rhythm of years to centuries. 
 However,  there could  be a significant loss of photons if the rays need to pass through much material such as fast moving clouds, or  being attenuated by the supermassive black hole silhouette.
\cite{1997ApJ...484..180S} showed,  in the case of  multiple black holes,   that the process underlying  the leading perturbation in the vicinity of the multiple black holes may not have a strictly periodic character. Bearing this in mind, the signal of such flare could have  quasi-periodic nature. 

The orbiting hot  or cold-spot model would be a natural explanation for the observed   light curves with flares and associated changes. However, the long term light curves are also  well described by a pure red noise, indicating statistical fluctuations in the accretion flow underlying the observed variability. As pointed in \cite{2017MNRAS.470.4112G}
spectroscopic and multiwavelength observations could settle  the debate. For example, data from 
 the Large Synoptic Survey Telescope (LSST) will allow more precise extraction of periodic signal from the light curve which can reveal additional perturbation more precisely. Even now,	a simple test of  extracting the sinusoidal signal  model from the present light curve of PG1302-102 showed  that the reminder of series  fluctuates
between -0.3 and 0.3 mag after MJD 56500. The shape of this feature is coherent and  resembles the gradient of  the inverted hump seen in the light curve. Other parts of  the reminder  fluctuate with magnitudes ten times or more smaller, and are negligible in comparison to the main fluctuation.
Upcoming  large surveys  will determine
distribution of   physical characteristics of flares and their periodic variation which will also help to test  the hot spot and red noise  models of light curves on larger sample of objects.

\section{Conclusion} \label{sec:con}
 We develop one possible physical model which could explain the variability of the optical flux and a slight perturbation of sinusoidal  feature of the optical light curve of PG1302-102  reported in \cite{2018ApJ...859L..12L}.
The dynamical properties of PG1302-102 are described by the model of   the orbital motion  in the SMBHB system and the  attenuation due to cold spot in  the accretion disk around the   more massive black hole. The model recovered orbital period of   1899 days.
Second,  the 2D correlation maps of oscillatory patterns in the observed and modeled light curve are 
determined  with of  our hybrid method for periodicity detection. The inferred periods are   $1972 \pm 254$ and $ 1873 \pm  250$ days in the observed and modeled light curves, respectively, which are slightly perturbed values in comparison to the \cite{2015Natur.518...74G} and close to the period predicted  by our physical model.
Our model suggests the perturbation  within the disk of  the  more massive component, in the form of a cold spot, as an explanation for  the  perturbed sinusoidal characteristic of the curve, which also slightly deformed  the detected period. Moreover,  our model gives  the  light curve a chance of  resembling   a sinusoidal variability within a few thousand days. Thus, future  monitoring of this object is important, and should  bring more light into dynamics of the object.

\
\

The authors thank to Suvi Gezari and Tingting Liu for providing the complete data set.
This work is supported by  the  project (176001)  \textit{Astrophysical
Spectroscopy of Extragalactic Objects} of  Ministry of Education, Science and Technological Development of Serbia.
 The authors would like to express gratitude to an anonymous Reviewer for
comments and suggestions which greatly  improved the quality of the
manuscript.




\begin{thebibliography}{}




\bibitem[Abramowicz et al.(1992)]{1992Natur.356...41A}
Abramowicz M. A., Lanza A., Spiegel E. A., Szuszkiewicz E., \ 1992, Nat, 356, 41 

 \bibitem[Abramowicz et al.(2010)]{2010A&A...521A..15A} 
 Abramowicz,  M. A., Jaroszy{\'n}ski, M.,   Kato, S.,   Lasota, J.-P,  R{\' o}z{\'a}{\'n}ska,  A.,   Sadowski  , A.,  \  2010, A\& A, 521, A15


 \bibitem[Abramowicz \& Fragile(2013)]{2013LRR....16....1A}  
Abramowicz, M. A., Fragile, P. C., \ 2013, Living Reviews in Relativity,  16,   id. 1


\bibitem[Angelakis et al.(2015)]{2015A&A...575A..55A}
Angelakis, E.; Fuhrmann, L.; Marchili, N.; Foschini, L.; Myserlis, I.; Karamanavis, V.; Komossa, S.; Blinov, D.; Krichbaum, T.P.; Sievers, A.; et al., \ 2015, A \& A, 2015, 575, A55




\bibitem[Armitage \& Reynolds(1992)]{2003MNRAS.341.1041A}
Armitage P. J., Reynolds C. S., \ 2003, MNRAS, 341, 1041


\bibitem[Balbus \& Hawley (1991))]{1991ApJ...376..214B}
Balbus S. A., Hawley J. F., \ 1991, ApJ, 376, 214


\bibitem[Bentz et al.(2007)]{2007ApJ...662..205B}
Bentz, M. C., Denney, K. D., Cackett, E. M.,  Dietrich, M., \ 2007, ApJ, 662, 205


\bibitem[Bon et al.(2012)]{2012ApJ...759..118B}
Bon, E.,  Jovanovi{\'c}, P., Marziani,  P.,  Shapovalova, A. I.,  Bon,  N., Borka Jovanovi{\'c}, V. et al. \ 2012, \apj, 759, 2, id.118 


\bibitem[Bon et al.(2016)]{bon16} Bon, E., Zucker, S., Netzer, H. et al. \ 2016,  \apjs, 225, 29

\bibitem[Bogdanovi{\'c} et al.(2008)]{2008ApJS..174..455B} 
Bogdanovi{\'c}, T., Smith, B. D., Sigurdsson, S.,  Eracleous, M., \ 2008, ApJS, 174, 455

\bibitem[Box \& Jenkins(1970)]{Box70}
Box, G. E. P.,  Jenkins, G. M. \ 1970,  Time Series Analysis, Forecasting and Control, San Francisco Holden Day


\bibitem[Bowen et al.(2018)]{2018ApJ...853L..17B}
Bowen, D. B., Mewes, V., Campanelli, M., Noble, S. C., Krolik, J. H., Zilh\~{a}o, M. \ 2018, \apjl, 853, id. L17


\bibitem[Britzen et al. (2018)]{2018MNRAS.tmp..975B}
Britzen, S., Fendt, C., Witzel, G., Qian, S.-J., Pashchenko, I. N. et al.  \ 2018,  accepted to \mnras 

\bibitem[Broderick \& Loeb  (2006)]{2006MNRAS.367..905B}
Broderick A. E., Loeb A., \ 2006, MNRAS, 367, 905 


\bibitem[Brockwell \&  Davis(2002)]{Brock02}Brockwell, P. J.,  Davis, R. A. \  2002,  Introduction to Time Series and Forecasting. New York: Springer.

 

\bibitem[Burnham \& Anderson(2002)]{Burn02}
Burnham, K.P.,  Anderson, D. R.,  \ 2002,  Model Selection and Multi-Model Inference A Practical Information-Theoretic Approach, Springer-Verlag, New York


\bibitem[Charisi et al.(2018)]{2018MNRAS.476.4617C}
Charisi, M.,  Haiman, Z.,  Schiminovich, D.,  D' Orazio, D. J. \ 2018, \mnras,  476, 4617


\bibitem[Dai et al.(2010)]{2010MNRAS.402.1614D}
Dai L., Fuerst S. V., Blandford R., 2010, MNRAS, 402, 1614


\bibitem[D' Orazio et al.(2013)]{2013MNRAS.436.2997D}	
D' Orazio, D. J., Haiman, Z.,  MacFadyen, A. 2013, MNRAS, 436, 2997


\bibitem[D' Orazio et al.(2015)]{2015Natur.525..351D}	
D' Orazio, D. J., Haiman, Z., Schiminovich, D. \ 2015, \nat, 525, 351

  
\bibitem[D' Orazio et al.(2016)]{2016MNRAS.459.2379D}	
D' Orazio, D. J., Haiman, Z., Duffell, P., MacFadyen, A.,  Farris, B. 2016, MNRAS, 459, 2379



\bibitem[D' Orazio \& Haiman(2017)]{2017MNRAS.470.1198D}
D' Orazio D. J.,  Haiman Z. \ 2017,\mnras, 470, 1198


\bibitem[D' Orazio \& Loeb(2017)]{2017arXiv171202362D}
D' Orazio, D. J., Loeb, A. 2017, sent to \apj, (arXiv:1712.02362)


\bibitem[Dov{\v c}iak et al. (2004)]{2004ApJS..153..205D}
Dov{\v c}iak, M., Karas, V., Yaqoob, T. \  2004, ApJS 153, 205  

\bibitem[Fageot et al. (2017)]{F17}
 Fageot, J.,  Uhlmann, V.,   Unser, M.,  \ 2017 Gaussian and Sparse Processes are Limits of Generalized Poisson Processes, submitted to Applied and Computational Harmonic Analysis, arXiv preprint,1702.05003


\bibitem[Farris et al.(2014)]{2014ApJ...783..134F}
Farris, B. D., Duffell, P., MacFadyen, A. I., Haiman, Z., \ 2014, ApJ, 783, 134

 

\bibitem[Fukue(2003)]{2003PASJ...55.1121F}
Fukue, J., \ 2003, PASJ, 55, 1121

 
\bibitem[Gaskell \& Harrington(2018)]{2018MNRAS.478.1660G}
Gaskell, C. M., Harrington, P. Z., \  2018, MNRAS,  478, 1660

\bibitem[Graham et al.(2015)]{2015Natur.518...74G}Graham, M. J.,  Djorgovski, S. G., Stern, D., Glikman, E.,   Drake, A. J. et al. \ 2015, \nat, 518, 74


\bibitem[Graham et al.(2017)]{2017MNRAS.470.4112G}
	Graham, M. J., Djorgovski, S. G., Drake, A.  J., Stern, D. et al. \ 2017, MNRAS, 470, 4112

\bibitem[Gould \& Miralda-Escud{\' e}(1997)]{1997ApJ...483L..13G}
Gould, A.,  Miralda-Escud{\' e}, J., \ 1997, ApJ, 483, L13


\bibitem[Jovanovi{\'c} et al.(2010)]{2010ApJ...718..168J}
Jovanovi{\'c}, P., Popovic{\'c}, L. {\v C}., Stalevski, M, Shapovalova, A. I., \ 2010, ApJ,  718, 168


\bibitem[Jun et al.(2015)]{2015ApJ...814L..12J}
Jun, H. D., Stern, D., Graham, M. J., Djorgovski, S. G., Mainzer, A., Cutri, R., M., Drake, A. J., Mahabal, A. A. \ 2015, \apjl, 814, L12


\bibitem[Karas \& Bao(1992)]{1992A&A...257..531K}
Karas V., Bao G. 1992, A\& A, 257, 531  

\bibitem[Karas (1999)]{1999PASJ...51..317K}
Karas V., 1999, PASJ, 51, 317 


\bibitem[Kasliwal et al.(2017)]{2017MNRAS.470.3027K}
Kasliwal, V. P., Vogeley, M. S.,  Richards, G. T., \ 2017, MNRAS, 470, 3027


\bibitem[Kaulakys \& Alaburda(2009)]{2009JSMTE..02..051K}
Kaulakys, B.,  Alaburda, M., \  2009,  Journal of Statistical Mechanics: Theory and Experiment,  02, 02051




\bibitem[Kelley et al.(2011)]{2011ApJ...730...52K}
Kelly, B. C., Sobolewska, M., Siemiginowska, A., \ 2011, ApJ, 730, 52 


\bibitem[Kelley et al.(2017)]{2017MNRAS.464.3131K}
Kelley, L. Z., Blecha, L.,  Hernquist, L. 2017, \mnras, 464, 3131


\bibitem[Khan et al.(2016)]{2016ApJ...828...73K}
 Khan, F. M., Fiacconi, D., Mayer, L., Berczik, P.,  Just, A. 2016, \apj, 828, 73

	

\bibitem[Kova{\v c}evi{\' c} et al.(2017)]{2017Ap&SS.362...31K}
Kova{\v c}evi{\' c}, A., Popovi{\' c}, L. {\v C}., Shapovalova, A. I., Ili{\' c}, D. \ 2017,  Ap\& SS, 362, id. 31





\bibitem[Kova{\v c}evi{\' c} et al.(2018)]{2018MNRAS.475.2051K}Kova{\v c}evi{\' c}, A.~B., P{\'e}rez-Hern{\'a}ndez, E., Popovi{\' c}, L. {\v C}., Shapovalova, A. I.,  Kollatschny, W.,  Ili{\' c}, D. \ 2018, \mnras, 475, 2051

\bibitem[Krolik \& Hawley (2002)]{2002ApJ...573..754K}
Krolik, J. H., Hawley, J. F., \ 2002, ApJ, 573, 754 

\bibitem[ Kudryavtseva et al.(2011)]{2011MNRAS.415.1631K}   
 Kudryavtseva, N.A., Gabuzda, D.C., Aller, M. F., Aller, H.D., \ 2011,   MNRAS,  415, 1631


\bibitem[Kwiatkowski et al.(1992)]{DOI: 10.1016/0304-4076(92)90104-Y}
Kwiatkowski, D., Phillips, P. C. B.,   Schmidt, P.,  Shin, Y. \ 1992, Journal of Econometrics, 54,   159


\bibitem[Lasota(2016)]{2016ASSL..440....1L}
Lasota, J.-P., \  2016, Astrophysics of Black Holes: From Fundamental Aspects to Latest Developments, Editor: Bambi, C.,  Chapter: Black Hole Accretion Discs, Springer Berlin Heidelberg, 1-60




 
\bibitem[Liu et al.(2007)]{2007ApJ...671..695L}
Liu, B. F, Taam, R. E, Meyer-Hofmeister, E., Meyer, F., \  2007,  ApJ, 671, 695




 
\bibitem[Liu et al.(2011)]{2011ApJ...726...10L}
Liu, B. F., Done, C., Taam, R. E., \  2011,  ApJ, 726, 10


\bibitem[Li et al.(2016)]{li16}
Li, Y.-R., Wang, J.-M., Ho, L. C., Lu, K.-X., Qiu, J., Du, P., Hu, C., 
Huang, Y.-K., Zhang, Z.-X., Wang, K., Bai, J.-M. \ 2016, \apj, 822, 4

\bibitem[Liu et al.(2018)]{2018ApJ...859L..12L}Liu,T.,  Gezari, S., Coleman Miller, M. \ 2018, \apjl,
 859,  id. L12
 
 \bibitem[Lobanov  \& Roland (2005)]{2005A&A...431..831L}
 Lobanov, A. P., Roland, J. \ 2005, \aap, 431,831
 

 \bibitem[MacFadyen \& Milosavljevi{\'c}(2008)]{2008ApJ...672...83M}
MacFadyen, A. I., Milosavljevi{\'c}, M., \ 2008, ApJ, 672,  83


 \bibitem[Mayer \& Pringle(2007)]{2007MNRAS.376..435M}
Mayer M, Pringle J. E.  \ 2007, MNRAS,  376, 435




\bibitem[Meyer et al.(2007)]{2007A&A...463....1M}
Meyer, F., Liu, B. F, Meyer-Hofmeister, E.,  \ 2007, A\& A,  463, 1





\bibitem[Meyer et al.(2009)]{2009A&A...508..329M}
Meyer-Hofmeister E., Liu B. F., Meyer F., \ 2009, A\& A, 508, 329


\bibitem[Nayakshin et al.(2004)]{2004A&A...413..173N}
Nayakshin S., Cuadra J., Sunyaev R., 2004, A\& A, 413, 173


 \bibitem[Noda(2015)]{Noda15}
Noda, I., \ 2015,  Biomedical Spectroscopy and Imaging,  4,  109 



\bibitem[Paczynski \& Rudak(1980)]{1980AcA....30..237P}
Paczynski, B., Rudak, B., \  1980, AcA, 30, 237

	



\bibitem[Palenzuela et al.(2010)]{2010Sci...329..927P}
Palenzuela, C., Lehner, L., Liebling, S. L.\ 2010, Science, 329, 927



\bibitem[Papaloizou \& Pringle(1977)]{1977MNRAS.181..441P}	
Papaloizou, J., Pringle, J. E., \ 1977, MNRAS, 181, 441



\bibitem[Pietil{\"{a}} (1998)]{1998ApJ...508..669P}
Pietil{\"{a}}, H., \ 1998,   ApJ, 508, 669 



\bibitem[Pech{\'a}{\v c}ek et al.(2008)]{2008A&A...487..815P}
Pech{\'a}{\v c}ek T., Karas V., Czerny B. \ 2008, A\& A, 487, 815


 \bibitem[Pech{\'a}{\v c}ek et al.(2013)]{2013A&A...556A..77P}
Pech{\'a}{\v c}ek, T., Goosmann, R. W., Karas, V., Czerny, B., Dov{\v c}iak, M.. \ 2013, A \& A,  556, id.A77


\bibitem[Popovi{\'c} (2012)]{2012NewAR..56...74P}Popovi{\'c}, L. {\v C}. \ 2012, \nar,  56, 2-3, 74

\bibitem[Popovi{\'c} et al.  (2012)]{2012A&A...538A.107P}
Popovi{\'c}, L. {\v C}. Jovanovi{\'c}, P., Stalevski, M., Anton, S., Andrei, A. H., Kova{\v c}evi{\'c}, J., Baes, M., \ 2012, A \& A,  538, id.A107

\bibitem[Popovi\'c \& Simi\'c(2018)]{Popovic18}Popovi\'c, L. \v C., \& Simi\'c, S. \ 2018, in preparation

\bibitem[Poutanen \& Fabian(1999)]{1999MNRAS.306L..31P}
Poutanen J., Fabian A. C., 1999, MNRAS, 306, L31



\bibitem[Press(1978)]{1978ComAp...7..103P}Press, W. ~H. \ 1978, Comments on Modern Physics, Part C - Comments on Astrophysics,  7,  103


\bibitem[Roedig et al.(2014)]{2014ApJ...785..115R}
Roedig, C., Krolik, J. H., Coleman, M. M., \ 2014, ApJ, 785,  115



\bibitem[R{\' o}{\. z}a{\' n}ska \& Czerny(2000)]{2000A&A...360.1170R}
R{\' o}{\. z}a{\' n}ska,  A., Czerny, B., \  2000,  A \& A,  360, 1170

\bibitem[Schnittman(2005)]{2005ApJ...621..940S}
Schnittman, J. D., \ 2005,  ApJ,  621, 940


\bibitem[Schnittman et al.  (2006)]{2006ApJ...651.1031S} 
Schnittman J. D., Krolik J. H., Hawley J. F. \  2006, ApJ, 651, 1031

\bibitem[Shahbaz (1999)]{1999JApA...20..197S}
Shahbaz, T. \ 1999, JA\& A, 20, 197
 
 \bibitem[Shakura \& Sunyaev(1973)]{Shakura73} Shakura, N.I., \& Sunyaev, R. A. 1973, A\& A, 24, 337

 
\bibitem[Shapovalova et al. (2008)]{2008A&A...486...99S}
Shapovalova, A. I.,  Popovi{\'c}, L. {\v C}., Collin, S.,  Burenkov, A. N., Chavushyan, V. H. et al.  \  2008,  A \& A,  486, 99


\bibitem[Shapovalova et al. (2010)]{2010A&A...517A..42S}
Shapovalova, A. I.,  Popovi{\'c}, L. {\v C}., Burenkov, A. N., Chavushyan, V. H., Ili{\'c}, D. et al.  \ 2010, A \& A, 517,  id. A42


\bibitem[Shapovalova et al. (2017)]{2017MNRAS.466.4759S}
Shapovalova, A. I., Popovi{\'c}, L. {\v C}., Chavushyan, V. H., Afanasiev, V. L., Ili{\'c}, D. et al.,  \ 2017, MNRAS, 466, 4759







\bibitem[Shi et al.(2012)]{2012ApJ...749..118S}
Shi, J.-M., Krolik, J. H., Lubow, S. H.,  Hawley, J. F. 2012, ApJ, 749, 118



\bibitem[Shi \& Krolik(2015)]{2015ApJ...807..131S}
Shi, J. M.,  Krolik, J. H.,  \ 2015, ApJ,  807,  id. 1

\bibitem[Simi\'c \& Popovi\'c (2016)]{Simic16}Simi\'c, S., \& Popovi\'c, L. \v C.  \ 2016, \apjs, 361, 59




\bibitem[Stella \& Vietri (1999)]{1999PhRvL..82...17S}
Stella, L., Vietri, M. \ 1999, Phys. Rev. Lett., 82, 17


\bibitem[Strubbe \&  Quataert (2009)]{2009MNRAS.400.2070S}
Strubbe, L. E.,   Quataert,  E., \ 2009,  MNRAS, 400,  2070




\bibitem[Sundelius et al.  (1997)]{1997ApJ...484..180S}
Sundelius, B., Wahde, M., Lehto, H. J.,Valtonen, M. J.,  \ 1997, ApJ, 484, 1,180


\bibitem[Tanaka (2012)]{2012EPJWC..3906008T}
Tanaka, T., \ 2012, Tidal Disruption Events and AGN Outbursts,  Edited by Saxton, R.,   Komossa, S., EPJ Web of Conferences, 39, id.06008
 
  
 \bibitem[Tanaka (2013)]{2013MNRAS.434.2275T}  
 Tanaka, T. L.,  \  2013, MNRAS,  434,2275
 
\bibitem[Tang et al.(2017)]{2017MNRAS.469.4258T}
Tang, Y., MacFadyen, A.,  Haiman, Z. , \ 2017, MNRAS, 469, 4258



\bibitem[Tang et al.(2018)]{2018MNRAS.476.2249T}
Tang, Y., Haiman, Z.,  MacFadyen, A., \ 2018, MNRAS,  476, 2249





\bibitem[Valtaoja et al. (1999)]{1999ApJS..120...95V}
Valtaoja, E., L{\"{a}}hteenm{\"{a}}ki, A., Ter{\"{a}}sranta, H., Lainela, M., \ 1999,  ApJS, 120, 95




\bibitem[Valtaoja et al.  (2000)]{2000ApJ...531..744V}   
Valtaoja, E., Ter{\"{a}}sranta, H., Tornikoski, et al., \  2000, ApJ, 531, 744




\bibitem[Vaughan (2010)]{2010MNRAS.402..307V} Vaughan, S.\ 2010, \mnras, 402, 307 




\bibitem[Vicente et al. (2014)]{Vicente14} Vicente, J. J., Mediavilla, E., Kochanek, C. S., Munoz, J. A., Motta, V., Falco, E, \& Mosquera, A. M. \ 2014, \apj, 783, 47


\bibitem[Yuan(2003)]{2003ApJ...594L..99Y}
Yuan,  F., \  2003, ApJ, 594, L99

\bibitem[Zamaninasab et al. (2010)]{2010A&A...510A...3Z}
Zamaninasab M., Witzel G., Eckart A., \ 2010, A\& A, 510, A3 



\bibitem[Zamaninasab et al. (2011)]{2011MNRAS.413..322Z}
 Zamaninasab, M.,  Eckart, A.,   Dov{\v c}iak, M.,   Karas, V. \ 2011,  MNRAS, 413, 322


\bibitem[{\. Z}ycki (2002)]{2002MNRAS.333..800Z}
{\. Z}ycki, P. T. \ 2002, MNRAS, 333, 800 






\end{thebibliography}
\end{document}